\documentclass[12pt]{article}
\usepackage{amssymb,amsmath,amsfonts}

\def\ba{\begin{array}}
\def\ea{\end{array}}

\def\ben{\begin{enumerate}}
\def\een{\end{enumerate}}
\def\beqan{\begin{eqnarray*}}
\def\eeqan{\end{eqnarray*}}
\def\btab{\begin{tabular}}
\def\etab{\end{tabular}}
\def\bit{\begin{itemize}}
\def\eit{\end{itemize}}

\newcommand{\lra}{\longrightarrow}
\newcommand{\cA}{{\cal A}}

\newcommand{\cZ}{{\cal Z}}

\newcommand{\cP}{{\cal P}}

\newcommand{\mbb}[1]{\mathbb{#1}}
\newcommand{\C}{{\mbb C}}
\newcommand{\CP}{{\C\mbb P}}
\newcommand{\Z}{{\mbb Z}}
\newcommand{\R}{{\mbb R}}

\begin{document}

\topmargin -2pt

\headheight 0pt

\topskip 0mm \addtolength{\baselineskip}{0.20\baselineskip}
\begin{flushright}
{\tt hep-th/0105265}
\end{flushright}

\vspace{10mm}

\begin{center}
{\large \bf   Noncommutative K3  Surfaces}\\

\vspace{7mm}

Hoil Kim\footnote{hikim@gauss.knu.ac.kr}\\

{\it Topology and Geometry Research Center, Kyungpook National University,\\
Taegu 702-701, Korea}\\

\vspace{3mm}

and \\

\vspace{3mm}

Chang-Yeong Lee\footnote{cylee@sejong.ac.kr}\\

{\it Department of Physics, Sejong University, Seoul 143-747, Korea}\\

\vspace{18mm}

\end{center}

\begin{center}
{\bf ABSTRACT}
\end{center}
We consider deformations of a toroidal orbifold $T^4/{\Z}_2$ and an
orbifold of quartic in $\CP^3$.
In the $T^4/{\Z}_2$ case,
we construct a family of noncommutative K3 surfaces
obtained via 
both complex and noncommutative deformations.
We do this following the line of algebraic deformation 
done by Berenstein and Leigh for the Calabi-Yau threefold.
We obtain 18 as the dimension of the moduli space both
in the noncommutative deformation as well as in the complex
deformation, matching the expectation from classical consideration.
In the quartic case, we construct a $4 \times 4$ matrix
representation of noncommutative K3 surface
in terms of quartic variables
in ${\CP}^3$ with a fourth root of unity.
In this case, the fractionation of branes occurs at
codimension two singularities due to the
presence of discrete torsion.
\\

\vfill

%\noindent
%PACS: 02.40.Gh, 11.25.Sq \\

\thispagestyle{empty}

%%%%%%%%%%%%%%%%%%%%%%%%%%%%%%%%%%%%%%%%%%%%%%%%%%%%%%%%%%%%%%%%%%%%%%%%%
\newpage
%%%%%%%%%%%%%%%%%%%%%%%%%%%%%%%%%%%%%%%%%%%%%%%%%%%%%%%%%%%%%%%%%%%%%%%%%
\section*{I. Introduction}

Starting from mid eighties, there have been some works on noncommutative manifolds,
notably on noncommutative torus \cite{cr,rief}.
Yang-Mills solutions on noncommutative two torus were described as isomorphic to 
the commutative underlying torus
by Connes and Rieffel \cite{cr} and the case for higher dimensional tori 
has been dealt with Rieffel \cite{rief}.

After the work of Connes, Douglas, and Schwarz \cite{cds}, noncommutative space
has been a focus of recent interest among high energy physicists in
relation with string/M theory.
The connections between string theory and noncommutative geometry \cite{conn} and
Yang-Mills theory on noncommutative space have been studied by many people \cite{sw}.
However, much of these works have been related with noncommutative tori,
the most well-known noncommutative manifold, or noncommutative ${\R}^n$.
Relevant to compactification,
noncommutative tori in particular have been the main focus \cite{nct,hv,t4ours}.
Among higher dimensional tori,
 Hoffman and Verlinde \cite{hv} first described the moduli space of noncommutative four torus
in the case of the projective flat connections from the viewpoint of physics,
and more general solutions were obtained in Ref. \cite{t4ours}.

So far, physically more interesting noncommutative version of
orbifolds or Calabi-Yau(CY) manifolds  have been seldom
addressed.  Until recently, only a few cases of ${\Z}_n$ type orbifolds of noncommutative tori
have been studied \cite{ks1,ks2,kkl}.
Orbifolds of four tori by a discrete group were studied by Konechny and Schwarz \cite{ks2}.
They determined
the K group of them.
In Ref. \cite{kkl}, projective modules on these $T^4_{\theta}$ were explicitly constructed
following the methods of
Rieffel \cite{rief}, then
 the dual structure of $Z_2$ orbifolds of them was considered.

Recently,  Berenstein, Jejjala, and Leigh\cite{bjl} initiated an algebraic geometry approach
to noncommutative moduli space.
Then,
 Berenstein and Leigh\cite{bl1} discussed noncommutative CY threefold from the
viewpoint of algebraic geometry. They considered two examples: a toroidal orbifold $T^6/{\Z}_2 \times
{\Z}_2$ and an orbifold of the quintic in ${\CP}^4$, each with
discrete torsion \cite{vafa,vfwt,doug,dgfl,jgomi}.
There, they explained the fractionation of branes at singularities
 from noncommutative geometric viewpoint, when discrete torsion is
present.
This case is different from the fractionation of branes due to the
resolution of singularities in the orbifolds without discrete torsion.
When discrete torsion is present, singularities cannot be resolved
by blow-up process \cite{vfwt}.
Orbifold singularities without discrete torsion
can be resolved by blow-ups.
In Ref. \cite{aspin95} the orbifold singularities in a K3 surface were  studied
and it was shown that the K3 moduli space is related with two form $B$-field.
The sixteen fixed points of $T^4/{\Z}_2$ can be blown-up to give a smooth K3 surface.
And among its 22-dimensional moduli space where two-form $B$-field lives,
16 components of $B$ come from the twisted sector due to orbifolding.
Then Douglas and company \cite{doug96,ddg} showed that this orbifold resolution
gives rise to fractional branes.

 Considering a D-brane world volume theory in the presence of
 discrete torsion,  Douglas \cite{doug} found that the resolution of singularities agrees with what
 Vafa and Witten \cite{vfwt} previously proposed.
 In this process, he found a new type of fractional branes bound to the singularities.
 Berenstein and Leigh \cite{bl1} sucessfully described this type of fractionation of branes
  from noncommutative
 geometric viewpoint, in response to the issue raised by Douglas \cite{doug}.
 Douglas pointed out that there had been no satisfactory
 understanding of discrete torsion from geometric viewpoint and
suggested noncommutative geometric
approach as a possible wayout.

In Ref. \cite{bl1},  Berenstein and Leigh first considered the $T^6/{\Z}_2 \times {\Z}_2$ case
 and recovered a large slice of the moduli space of complex structures of
 the CY threefold
 from the deformation of the noncommutative resolution of the orbifolds
via central extension of the local algebra of holomorphic
 functions.
Then, they considered  the orbifolds of the quintic Calabi-Yau threefolds
and constructed an explicit representation of a family
of the noncommutative algebra.

 Here, we apply this algebraic approach to K3 surfaces
 in the cases of the orbifolds $T^4/{\Z}_2$ and the orbifolds of the quartics in ${\CP}^3$.
In the first example, we did in two steps.
As a preliminary step, we follow Berenstein and Leigh \cite{bl1}
obtaining a similar result.
We deformed
$E_1 \times E_2$ yielding the two dimensional complex
deformation of Kummer surface whose moduli space is
of dimension 2.
As a main step, we construct a family of
noncommutative K3 surfaces by algebraically deforming  $T^4/{\Z}_2$
in both complex and noncommutative ways at the same time.
Our construction shows that the dimensions of moduli space
for both the complex structures and the noncommutative deformation
are the same, 18. And this is the dimension of the moduli space
of the complex structures of K3 surfaces constructed with two elliptic
curves.
In the commutative K3 case, the moduli space for the K3 space itself has been known already
 (see for instance \cite{aspinwall}),
and even the moduli space for the bundles on K3 surfaces
 has been studied \cite{mukai}.
In the first example of  CY threefold case \cite{bl1},  the three holomorphic coordinates $y_i$
anticommute with each other to be compatible with ${\Z}_2$ discrete torsion.
Thus, in this case the central extension of the local algebra can be understood as
 possible deformations of $su(2)$ in the noncommutative direction for the
 underlying 6 torus, but the  deformations of the CY threefold were
  done via the newly obtained center from the central extension by which
  most of the moduli space of complex deformations was obtained.
However, in our case we obtained the same dimension of the moduli spaces
both in the complex deformation and in the noncommutative deformation.

In the case of the orbifold of the quartic in ${\CP}^3$, we obtain the result similiar to
 the quintic threefold case:
We see the fractionation of branes at codimesion two singularities instead of codimension
three singularities. And this case corresponds to a generalization of
 $su(2)$ deformation.

Our main methods are noncommutative algebraic geometry after
 Refs. \cite{bl1} and \cite{bjl}.
In  Refs.\cite{bl1,bjl},
 the algebra of holomorphic functions on a non-commutative algebraic space
was used 
instead of (pre) $C^*$ algebra of (smooth) continuous functions.
Then the center of the algebra is the commutative part
 inside the noncommutative algebra.
They correspond to two geometries:
a commutative geometry on which closed strings propogate, and a
noncommutative version for open strings.
The center of the algebra describes locally a
commutative Calabi-Yau manifold with possible singularities.
Here, our approach is a little different from theirs.
We deform the invariant polynomials of the K3 surface itself,
such that its center is ${\mbb P}^1 \times {\mbb P}^1$
not the classical K3 surface itself, unlike the Berenstein and Leigh's
case \cite{bl1,bjl}.

In section II, we construct a family of noncommutative
 orbifolds of $T^4/{\Z}_2$, and
obtain the moduli
space of noncommutative K3 (Kummer) surfaces.
In section III, we deal with the orbifold of quartic surfaces in ${\CP}^3$.
Here, we construct a four dimensional representation of a family of noncommutative algebra.
In section IV, we conclude with discussion.

%\pagebreak

%%%%%%%%%%%%%%%%%%%%%%%%%%%%%%%%%%%%%%%%%%%%%%%%%%%%%%%%%%%%%%%%%
\section*{II. Orbifold of the torus}\label{t4z2}

In this section, we
 consider the orbifold of $T^4/{\Z}_2$. Here we will use the compact orbifold as a target space for
D-branes and
 consider $T^4$ as the product of two elliptic curves, each given in the Weierstrass form
\begin{equation}
\label{t2}
y_i^2 = x_i (x_i-1)(x_i - a_i)
\end{equation}
with a point added at infinity for $i=1,2.$
The ${\Z}_2$ will act by $y_i \lra \pm y_i$ and $x_i \lra x_i$ so that the holomorphic
function $y_1 y_2$
%
% and the holomorphic two form $dy_1 \wedge dy_2$
%
is invariant under the
orbifold action.
This satisfies the CY condition on the quotient space. The four fixed
points of the orbifold at each torus are located at $y_i =0$ and at the point of infinity.
By the following change of variables, the point at infinity is brought to a finite point:
\begin{eqnarray}
\label{t2tr}
y_i \lra y_i' = \frac{y_i}{x_i^2}, \\
x_i \lra x_i' = \frac{1}{x_i} . \nonumber
\end{eqnarray}
It is known that $H^2(Z_m \times Z_n,U(1))=Z_d$,
where $d$ is the greatest common divisor of $m$ and $n$ \cite{kpsky, jgomi}.
So, in this case there is no discrete torsion.
 In the CY threefold case,
$T^6/{\Z}_2 \times {\Z}_2$ has discrete torsion, and the presence of discrete torsion requires
that $y_i$ ($i=1,2,3$) for three elliptic curves be anti-commuting variables, where these variables
can be represented with $su(2)$ generators.

In the $T^4/{\Z}_2$ case,
we consider the quotient space by the invariant polynomials.
They are $x_1, x_2$ and $y_1y_2=t$ with the constraints
$t^2=f_1(x_1)f_2(x_2)$ coming from  $y_1^2=f_1(x_1)$ and
$y_2^2=f_2(x_2)$.
This is the singular Kummer surface doubly covering ${\mbb P}^1 \times
{\mbb P}^1$ with 4 parallel lines meeting 4 parallel lines once as branch
locus. These are just the locus of zeroes of $f_1(x_1)f_2(x_2)$,
where $f_i$ is considered to be of degree 4 in ${\mbb P}^1$ instead of
${\C}$ including the infinity.

In our apparoach for noncommutative K3 surface,
we do in two steps.
 As a preliminary step, following the line
of Berenstein and Leigh \cite{bl1}
we deform the covering space such that the center
of the deformed algebra corresponds to the commutative K3 surface.
In this process, the complex structure of the center is also deformed 
as a consequence of the covering space deformation.
As a main step, 
we deform the K3 (Kummer) surface itself in the noncommutative direction.

In the first step, we consider two variables for the
classical variable $t=y_1y_2$,
namely $t_1$ for $y_1y_2$ and  $t_2$ for $y_2y_1$.
In the classical(commutative) case,
both $y_1y_2$ and $y_2y_1$ are invariant
under the $Z_2$ action, and they are the same.
Now, we consider a variation of four torus with
$ y_1 y_2 + y_2 y_1 =0 $.
Then the invariant polynomials under ${\Z}_2$
action are generated by
 $ x_1, x_2, t_1=y_1 y_2 = -y_2 y_1 = -t_2$
and they satisfy $t_1^2=t_2^2 =-f_1 f_2 $.
Now, we deform it into
\begin{equation}
\label{ctyP}
y_1y_2  + y_2y_1 = P_0(x_1,  x_2)
\end{equation}
where $P_0$ is a polynomial of degree two for each variable \cite{bl1}.
Then the subalgebra generated by invariant polynomials
($x_1, x_2, t_1, t_2 = -t_1 + P_0$)
is not the center but is a commuting subalgebra of
the deformed algebra.
They satisfy the condition $t_i^2 -P_0t_i +f_1f_2=0,$
which can be rewritten as
\begin{equation}
\label{cpdf}
(t_i - \frac{P_0}{2})^2 = \frac{P_0^2}{4} -f_1 f_2, ~~ {\rm for}~~ i=1,2.
\end{equation}
Since this is a commuting subalgebra, it has a geometrical meaning.
It is a K3 surface doubly covering ${\mbb P}^1 \times {\mbb P}^1 $ with
branch locus $ \frac{P_0^2}{4} -f_1 f_2 =0. $
Note that this is a polynomial of degree four for each $x_i, ~~ i=1,2$.
The noncommutative deformation of $E_1 \times E_2 $
corresponding to the condition (\ref{ctyP}) induces a
complex deformation of K3 surface corresponding to the condition
(\ref{cpdf}).
 Now, we count the parameters of $P_0(x_1, x_2)$. It 
has 9 parameters since it is of degree two  in each variable.
But we must subtract one corresponding to multiplication by a constant
and six coming from  $PGL(2,{\C})$  action on both
variables $x_1, x_2$. So, the number of the
 remaining degrees of freedom is two.
Actually, both the dimension of the K\"{a}hler metric spaces
in  $E_1 \times E_2 $  and the deformation dimension of
the Kummer surface coming from   $E_1 \times E_2 $
are 2.

As a main step, we now consider the noncommutative deformation of K3 surface itself.
 From the previous consideration of commutative K3 surface, we may 
 consider the case with four variables, $x_1, x_2, t_1,
t_2$. This is simply
 because that there is no apriori reason that $t_1$ and $t_2$
 be the same in the noncommutative case though they are the same
in the classical case.
Thus we begin with four independent variables for the noncommutative
$T^4/{\Z}_2$.
However, we do not know yet how to proceed to the noncommutative
case in which all the variables are non-commuting each other.
So, we consider a simpler case in which
$x_1$, $x_2$ are commuting variables
and belong to the center and only $t_1$, $t_2$ are non-commuting variables.

To proceed in that direction, we first consider the cases of
the independent complex
deformation for each $t_i$($ i=1,2$).
Let us consider the case with $t_1$ but not with $t_2$
in which we
 deform the relation $t_1^2=f_1(x_1)f_2(x_2)$ to any
polynomial $h_1(x_1,x_2)$ of degree 4 in each variable. Then, for a
generic $h_1$ it determines a smooth K3 surface doubly covering ${\mbb P}^1
\times {\mbb P}^1$. If we count the number of deformation parameters naively,
we have 25 parameters since $h_1$ is a polynomial of degree 4
in each variable.
However, we must ignore 1 parameter since multiplication by a constant
defines an isomorphic K3 surface with exactly the same branch locus.
We should also consider another symmetry giving isomorphic surfaces.
These are $PGL(2,{\C})$ acting both ${\mbb P}^1$ independently. Thus we have to
subtract 6 parameters, and we are left with 18 parameters. This is what
we expected since the moduli space of K3 surfaces doubly covering
${\mbb P}^1 \times {\mbb P}^1$ is of dimension 18.
This is the classical(complex)
deformation of the original singular Kummer surface.

We can consider in the same manner with $t_2$,
by deforming $t_2^2=f_1(x_1)f_2(x_2)$ to any
polynomial $h_2(x_1,x_2)$ of degree 4 in each variable.
Thus, this case also represents another complex deformation of the
Kummer surface.

When we have both $t_1$ and $t_2$, and  $h_1(x_1,x_2)\neq h_2(x_1,x_2)$,
the following two cases are seemingly possible.
\begin{equation}
\label{center}
[t_1, t_2]_{\mp} = c (\cZ \cA)
\end{equation}
Here, $c (\cZ \cA)$ is a function in the center of the local algebra, and
 ``$-$" denotes commutator and ``+" denotes anticommutator.
However, our requirement that $t_i^2$($i=1,2$) belong to the center,
namely $[t_1^2, t_2^2]=0$, allows only the anticommutator case.

Since only two variables $x_1, \ x_2$ are in the center of the
deformed algebra, the right hand side of (\ref{center}) should
be a polynomial and free of poles in each patch.
Thus,  (\ref{center}) can be written in the following form
\begin{equation}
\label{ctrP}
t_1t_2  + t_2t_1 = P(x_1,  x_2)
\end{equation}
where $P$ is a polynomial.
The change of variables (\ref{t2tr}) on $x_i$, now changes into
\begin{eqnarray}
\label{nctr}
t_j \lra t_j' = \frac{t_j}{x_i^2}, \ \ {\rm for} \ \ j=1 \ \ {\rm and} \ \ 2,  \\
x_i \lra x_i' = \frac{1}{x_i}, \ \ {\rm for} \ \ i=1 \ \ {\rm or} \ \ 2, \nonumber
\end{eqnarray}
and thus $P$ transforms as
\begin{equation}
\label{p12tr}
P(x_1, x_2) \lra x_1^4 P(1/x_1, x_2),
\end{equation}
under the transform of $x_1$.
Here, $P$ should be of degree at most four in $x_1$,
since it has to transform into
a polynomial. In a similar manner, it is
easy to see that it is of degree at most four in $x_2$.

Now, we have a noncommutative K3 surface defined by $x_1, x_2, t_1, t_2$
whose  noncommutativity is characterized by the relation (\ref{ctrP}),
 $t_1t_2+t_2t_1=P(x_1, x_2)$.

This is a deformation in another direction, a noncommutative deformation.
Here, we also have 25 parameters by naive counting.
But, due to the same
reason as in the classical(complex) deformation case described before,
 we have only 18 parameters.

In order to understand the above obtained moduli space of noncommutative deformations,
we first need to understand the algebraic and complex structures in the
commutative case.
 The aspect in the commutative case can be understood by looking into $(p,q)$
 forms preserved under the involution. In the present case,
  $(p,q)=(0,0),(1,1),(2,2),(2,0),(0,2)$ are preserved
under the involution. The dimension of preserved $(p,q)$ forms is
one except the case of $(1,1)$. The dimension of preserved $(1,1)$
forms is four as in the four torus. The rank of the Picard group
which is the intersection of $(1,1)$ forms with the integral
second cohomology class of the quotient space is at least two
since we constructed the four torus with two elliptic curves.
If we resolve singularities of the K3 surface,
 we have a two dimensional family of K3 surfaces whose Picard rank is at
least 18, since we have additional 16 coming from 16 singular
points. This corresponds to the so-called A model \cite{vfwt}.
Since there is no torsion we can deform the singularities, and get
a family of smooth K3 surfaces of dimension 18 described as above.
Each member of this family has the Picard group containing those
of the quotient spaces. This corresponds to the so-called B model
\cite{vfwt}.
 Note that a K3 surface in the family we are considering is not generic
since we started at a torus which is a product of two elliptic
curves, not a generic torus.
Vafa and Witten \cite{vfwt} considered the classical deformation of a
Calabi-Yau threefold doubly covering ${\mbb P}^1 \times {\mbb P}^1
\times {\mbb P}^1$.
Berenstein and Leigh \cite{bl1} considered the anticommutation of the
variables $y_i$, while $x_i$ are still commuting. Then they
considered the center generated by four variables
$x_1,x_2,x_3,w=y_1y_2y_3$ with the constraints
$w^2=f_1(x_1)f_2(x_2)f_3(x_3)$ corresponding to Calabi-Yau
threefolds. They deformed the 6 torus with anticommuting $y_i$ and
commuting $x_i$ as $y_iy_j+y_jy_i=P_{ij}(x_i,x_j)$ giving rise to
the change of the center into another center generated by
$x_1,x_2,x_3,w'=y_1y_2y_3+P_{13}y_2-P_{23}y_1-P_{12}y_3$ with the
constraints ${w'}^2=f_1f_2f_3+h(x_1,x_2,x_3)$, where the function
$h$ is determined by $P_{12},P_{13},P_{23}$. Notice that here $y_i$ 
are not  variables for Calabi-Yau threefold since
they are not invariant under the action. So, the deformation of
anticommuting variables affects the deformation of $T^6$ and the
central Calabi-Yau threefold at the same time.

In our present case,
 $t_1$ for $y_1y_2$ and $t_2$ for $y_2y_1$ are all invariants of the K3
surface.
In the preliminary step, we deformed $E_1 \times E_2 $ noncommutatively
giving rise to the two dimensional
 classical deformation of Kummer surfaces corresponding
to the invariant polynomials which is not the center but is a commuting
subalgebra. This is in line with Ref.\cite{bl1} in the sense that
noncommutative deformation of covering space yields complex deformation
of the classical space which corresponds to the center of the deformed
algebra.
In the main step, we deformed the classical(commutative) space itself
rather than the covering space.
As a result the center in this case does not correspond to the classical
space, and the deformation comprises both noncommutative
deformation with anticommuting $t_1,t_2$
and complex deformation arising from the constraints on $t_i^2, ~~ i=1,2$.
The dimensions of the moduli spaces of these deformations are 18
for both the noncommutative and complex deformation cases.

Our result in this section that 
the moduli space of the complex
structures of the commutative K3 surfaces and  the moduli space of
the noncommutative K3 surfaces have the same dimension 18 seemingly
suggests that there may be another kind of mirror symmetry in
these deformations.
However, since our deformation in the noncommutative direction
is not as general as it can be, we leave this as an open question.

%\pagebreak

%%%%%%%%%%%%%%%%%%%%%%%%%%%%%%%%%%%%%%%%%%%%%%%%%%%%%%%%%%%%%%
\section*{III. Orbifold of the quartic }\label{cp3}

In this section we consider a noncommutative version of
an orbifold of the quartic in ${\CP}^3$.
The complex structure moduli space of the quartic
has a discrete symmetry group
 of $\Z_4 \times \Z_4$.
The quartic  is described by
\begin{equation}
\label{quartic}
{\cP}(z_j) = z_1^4 + z_2^4 + z_3^4 + z_4^4 +\lambda z_1 z_2 z_3 z_4 = 0
\end{equation}
where the $z_j$ ( $ j=1, ..,4$ ) are homogeneous coordinates on ${\CP}^3$.
The $\Z_4 \times \Z_4$ action is generated by phases acting on the $z_j$ as
$z_j \lra w^{a_j} z_j ,$ with $w^4 =1,$ and the vectors
\begin{eqnarray}
\label{a-vector}
\vec{a}_1 = (1,-1,0,0) \nonumber \\
\vec{a}_2 = (1,0,-1,0)
\end{eqnarray}
consistent with the CY condition $\sum_{j=1}^4 a_j =0 $ mod 4.
Once we choose the action such that $z_4$ is invariant, then we can  consider a coordinate patch
where $z_4 =1$.
Here, the discrete torsion is given by \cite{kpsky,doug}
\begin{equation}
\label{dsctorsion}
H^2(\Z_4^2, \ U(1) ) = \Z_4
\end{equation}
so we need a phase to determine the geometry.

Within the coordinate patch, $z_i^4$ and $z_1 z_2 z_3$ are invariant.
We assume the following commutation relations among $z_j$
\begin{eqnarray}
\label{cp3cr}
z_1 z_2 = \alpha z_2 z_1 \nonumber \\
z_1 z_3 = \beta z_3 z_1 \nonumber \\
z_2 z_3 = \gamma z_3 z_2 ,
\end{eqnarray}
 and
 require that the above invariant quantities remain in the center of the noncommutative quotient
 space.
  From this requirement, one can easily see that
 \begin{equation}
 \alpha^4 =1,  \ \  \beta =\alpha^{-1}, \  \  \gamma = \alpha .
 \end{equation}

%%%%%%%%%%%%%%%

Now, we consider a four-dimensional representation of this with
the following matrices
\begin{equation}
\label{matrices}
P =  \hbox{diag}(1,\alpha,\alpha^2,\alpha^3),
% \ \ \   Q = \left(  \begin{array}{cccc}  0 & 0 &0  &1\\
% 1 &0 &0  &0\\
% 0 & 1 & 0 &0\\
% 0 & 0  &1 &0
% \end{array} \right)
\ \ \
Q = \begin{pmatrix} 0 & 0 &0  &1\\
1 &0 &0  &0\\
0 & 1 & 0 &0\\
0 & 0  &1   & 0
\end{pmatrix}
\end{equation}
where $\alpha$ is a fourth root of unity.
In terms of $P, \ Q $ matrices, we  put
$ z_1 = b_1 P,$  $z_2 = b_2 Q$, and $z_3 =b_3 P^m Q^n$
where $b_j$
are arbitrary complex numbers. The integers $m,n$ can be determined from the
requirement that $z_1 z_2 z_3 z_4$ belongs to the center of the algebra,
and are given by $m=n=3$.
By the use of the commutation relation $PQ = \alpha QP$ and of $P^4=Q^4=1$,
the defining relation of the quartic (\ref{quartic}) now becomes
\begin{equation}
\label{quartic-rep}
 b_1^4 +  b_2^4  + \alpha^2  b_3^4 + \alpha \lambda  b_1   b_2  b_3 = -1 .
\end{equation}
Notice that only
when $\alpha =1$, (\ref{quartic-rep}) satisfies the same classical relation (\ref{quartic}).
When the phase $\alpha$ becomes $-1$ which is a fourth root of unity,
the $z_j$'s become to anticommute, (\ref{cp3cr}). In this case
the algebra may be regarded as a deformation of $su(2)$.
Indeed we have
\begin{equation}
\label{algebraic rel}
z_iz_j=g_{ijk}z_k^{-1},
\end{equation}
where $g_{ijk}$ are totally antisymmetric scalars. Note that in $su(2)$,
we have $z_iz_j=if_{ijk}z_k=if_{ijk}z_k^{-1}$,
where we considered $su(2)$ as an algebra, not as a Lie algebra.
This is in line with
$T^6/{\Z}_2 \times {\Z}_2$ representation of the CY threefold with discrete torsion
in Ref. \cite{bl1} where
algebras  generated by three anticommuting variables
were considered as deformations of $su(2)$.

%%%%%%%%%%%%%%

If we think of the trace of the representation matrices as functions on the $b_i$ space
with the defining equation above, then the representation becomes reducible only
when two out of the three $z_i$ act
by zero, which shows that there are fractional branes for the codimension two
singularities.
In the case of the quintic threefold, the representation becomes not reducible
on the codimension two singularities.
There the representation becomes reducible on the codimension three
singularities where three out of the four $z_i$ act
by zero.
One may consider the coordinate transformations to other patches. It is not difficult to show that
one can still do the standard coordinate changes of the quartic. One needs to be only careful
with orderings of variables. In every coordinate patch the algebra is of the same type as (\ref{cp3cr})
as in the quintic case.
%%%%%%%%%%%%%%

%\pagebreak

%%%%%%%%%%%%%%%%%%
\section*{IV. Discussion}

In the first part of this paper, we studied
the deformations of
orbifolds of four tori which are products of two elliptic curves.

As a preliminary step, we followed Berenstein and Leigh \cite{bl1}
obtaining a similar result. We deformed $E_1 \times E_2$ yielding
the two dimensional complex deformation of the Kummer surface.
This agrees with the fact that the moduli space of the
Kummer surface coming from the product of two elliptic curves
is of dimension 2.

Then we deformed the singular K3 (Kummer) surface
in both noncommutative and complex directions.
We described the noncommutative K3 surfaces with
the two non-commuting variables coming from one of the classically
invariant variables. The squares of each of these two non-commuting
variables  belong to the center and 
 represent complex deformations of the Kummer surfaces.
Allowed deformations of the commutation relation for these
 two variables represent the noncommutative deformations.
This result is different from the
noncommutative deformation based on the 
${\Z}_2$ orbifold of
 noncommutative four torus $T^4_{\theta}$
 which has been carried out previously \cite{kkl}.
Here, we obtained 18 as the dimension of the moduli space of
our noncommutative deformations.
In the commutative case, 
the complete family of commutative K3 surfaces is of 20 dimension
inside which algebraic K3 surfaces are of 19 dimension.
However, the family coming from the ${\Z}_2$
quotient of $T^2 \times T^2$ is of 18 dimension.
This shows that the dimension of the moduli space of our noncommtative
deformations is the same as the one we get from the classical consideration
of complex deformations.

Notice that what we have done here is a little
different from the work of Berenstein and Leigh
\cite{bl1}.
In their work
they deformed the algebra of the covering space, in which
 the center of the deformed algebra corresponds to the classical
 commutative space of its undeformed algebra. 
On the other hand,
we deformed the K3 surface itself rather than
deforming the algebra of the covering space.
As a consequence,
its center is ${\mbb P}^1 \times {\mbb P}^1$ not the commutative
K3 surface.
           
In the second part of the paper,
we dealt with the orbifolds of the quartic K3 surfaces. In this case,
we have nonvanishing discrete torsion.
This turned out to be a deformation of $su(2)$  as in the
$T^6/{\Z}_2 \times {\Z}_2$ case of Ref. \cite{bl1}.
We could find fractional branes on the codimension two singularities
rather than on the codimension
three singularities as in the case of the quintic threefolds \cite{bl1}.
However, the reasoning was the same as in the quintic case.

We described  almost all the
complete family of K3 surfaces, but this does not fit to the most
general algebraic K3 surfaces of 19 dimensional family.
One should also find the suitable representation of the
noncommutative algebra we used, so that
one can find the dual noncommutative K3 and the symmetry group.
 Finally, from our result of the first part,
one may speculate a kind  of
mirror symmetry between complex deformations and noncommutative
deformations. Here, we will leave it as an open issue.

%\pagebreak

\vspace{5mm}
\noindent
%%%%%%%%%%%%%%%%%%%%%%%%%%
{\Large \bf Acknowledgments}

\vspace{5mm}
%\noindent
This work was supported by
KOSEF Interdisciplinary Research Grant No. 2000-2-11200-001-5.
We would like to thank Igor Dolgachev for useful discussions.

%\pagebreak

%%%%%%%%%% Macro for References %%%%%%%%%%%%%%%%%%%%%%%%%%%%%%%%

\newcommand{\MPL}{Mod.\ Phys.\ Lett.}
\newcommand{\NP}{Nucl.\ Phys.}
\newcommand{\PL}{Phys.\ Lett.}
\newcommand{\PR}{Phys.\ Rev.}
\newcommand{\PRL}{Phys.\ Rev.\ Lett.}
\newcommand{\CMP}{Commun.\ Math.\ Phys.}
\newcommand{\JMP}{J.\ Math.\ Phys.}
\newcommand{\JHEP}{JHEP}
\newcommand{\ib}{{\it ibid.}}

%%%%%%%%%%%%%%%%%%%%%%%%%%%%%%%%%%%%%%%%%%%%%%%%%%%%%%%%%%%%%%%%%%%
%%%%%%%%%%%%%%%%%%%%%%%%%%%%%%%%%%%%%%%%%%%%%%%%%%%%%%%%%%%%%%%%%%%%%%%%
%\newpage

\end{document}